\documentclass[a4paper,11pt]{article}
\usepackage{jcappub} % for details on the use of the package, please see the JINST-author-manual
\usepackage{lineno}
\usepackage{pifont}% http://ctan.org/pkg/pifont
\usepackage{array}
\usepackage{siunitx}
\usepackage{multirow}
\usepackage{graphicx}
\usepackage{subfig}
\usepackage{multicol}
\usepackage{comment}
\usepackage{dsfont}
\usepackage{soul}
\usepackage{txfonts}
\usepackage{color}
\usepackage[numbers]{natbib}

\arxivnumber{arXiv:2409.18714} % Only if you have one
\title{\boldmath Spectral Imaging with QUBIC: building frequency maps from Time-Ordered-Data using Bolometric Interferometry}

\author[a]{M.~Regnier,}
\author[a]{T.~Laclavere,}
\author[a]{J-Ch.~Hamilton,}
\author[b]{E.~Bunn,}
\author[a, c]{V.~Chabirand,}
\author[a]{P.~Chanial,}
\author[a]{L.~Goetz,}
\author[a]{L.~Kardum,}
\author[a]{A.~Huchet,}
\author[a]{P.~Masson,}
\author[d]{N.~Miron Granese,}
\author[e]{C.G.~Scóccola,}
\author[a, f]{S.A.~Torchinsky,}
\author[g, h]{E.~Battistelli,}
\author[i, j]{M.~Bersanelli,}
\author[g, h]{F.~Columbro,}
\author[g, h]{A.~Coppolecchia,}
\author[d]{B.~Costanza,}
\author[g, h]{P.~De Bernardis,}
\author[g, h]{G.~De Gasperis,}
\author[g, h]{S.~Ferazzoli,}
\author[k]{A.~Flood,}
\author[a]{K.~Ganga,}
\author[l, m]{M.~Gervasi,}
\author[a]{L.~Grandsire,}
\author[i, j]{E~.Manzan,}
\author[g, h]{S.~Masi,}
\author[i, j]{A.~Mennella,}
\author[n]{L.~Mousset,}
\author[k]{C.~O'Sullivan,}
\author[g, h]{A.~Paiella,}
\author[g, h]{F.~Piacentini,}
\author[a]{M.~Piat,}
\author[o]{L.~Piccirillo,}
\author[p]{E.~Rasztocky,}
\author[q, r]{M.~Stolpovskiy,}
\author[l, m]{M.~Zannoni}
\affiliation[a]{Astroparticule et Cosmologie, Université Paris Cité, CNRS, F-75013 Paris, France}
\affiliation[b]{University of Richmond, Richmond, USA}
\affiliation[c]{Ecole polytechnique, Institut Polytechnique de Paris, Palaiseau, France}
\affiliation[d]{Facultad de Ciencias Astronómicas y Geofísicas (Universidad Nacional de La Plata), Argentina}
\affiliation[e]{Cosmology and Theoretical Astrophysics group, Physics Department, FCFM, Universidad de Chile, Blanco Encalada 2008, Santiago, Chile}
\affiliation[f]{Observatoire de Paris, Université PSL, F-75013 Paris, France}
\affiliation[g]{Università di Roma - La Sapienza, Italy}
\affiliation[h]{INFN sezione di Roma, 00185 Roma, Italy}
\affiliation[i]{Universita degli studi di Milano, Italy}
\affiliation[j]{INFN sezione di Milano, 20133 Milano, Italy}
\affiliation[k]{National University of Ireland, Maynooth, Ireland}
\affiliation[l]{Università di Milano - Bicocca, Italy}
\affiliation[m]{INFN Milano-Bicocca, Italy}
\affiliation[n]{Laboratoire de Physique de l’École Normale Supérieure, ENS, Univ. PSL, CNRS, Sorbonne Univ., Univ. Par Cité, 75005 Paris, France}
\affiliation[o]{University of Manchester, UK}
\affiliation[p]{Instituto Argentino de Radioastronomía (CONICET, CIC), Argentina}
\affiliation[q]{International Space Science Institute (ISSI), Hallerstrasse 6, 8012 Bern, Bern, Switzerland}
\affiliation[r]{University of Bern, Hochschulstrasse 4, 3012 Bern, Switzerland}

\emailAdd{regnier@apc.in2p3.fr}

\abstract{{The search for relics from the inflation era in the form of B-mode polarization of the CMB is a major challenge in cosmology. The main obstacle appears to come from the complexity of Galactic foregrounds that need to be removed. Multi-frequency
observations are key to mitigating their contamination and mapping primordial fluctuations.}
   {We present ``Spectral-Imaging'', a method to reconstruct sub-frequency maps of the CMB polarization within the instrument’s physical bandwidth, a unique feature of Bolometric Interferometry that could be crucial for foreground mitigation as it provides an increased spectral resolution.}
   {Our technique uses the frequency evolution of the shape of the Bolometric Interferometer’s synthesized beam to reconstruct frequency information from the time domain data. We reconstruct sub-frequency maps using an inverse problem approach based on detailed modeling of the instrument acquisition. We use external data to regularize the convergence of the estimator and account for bandpass mismatch and varying angular resolution.}
   {The reconstructed maps are unbiased and allow exploiting the spectral-imaging capacity of QUBIC. Using end-to-end simulations of the QUBIC instrument, we perform a cross-spectra analysis to extract a forecast on the tensor-to-scalar ratio constraint of $\sigma(r)= 0.0225$ after component separation.}}

\begin{document}
\maketitle
\flushbottom

\section{Introduction}
    
    The current standard cosmological model assumes a primordial phase of accelerated expansion, called inflation, that provides initial conditions in good agreement with observational data~\citep{Ade_2021}. Searching for direct evidence of inflation is one of the main challenges in observational cosmology. Besides the observed scalar perturbations, inflation predicts the production of primordial gravitational waves, or tensor perturbations, that distinguishes between various inflationary models~\citep{pieroni:tel-01393860}. Tensor perturbations will leave a characteristic imprint in the Cosmic Microwave Background (CMB), contributing to a small amount of temperature and E-mode polarization anisotropies, and are the only known source of primordial B-mode anisotropies. The ratio between tensor and scalar power spectra is defined as the tensor-to-scalar ratio $r$. The detection of primordial B-modes, and therefore a non-zero value of $r$, would represent a major step forward in understanding the history of our Universe, particularly in context of inflationary theories. The current best 95\% confidence level (C.L.) upper-limit is $r<0.032$~\citep{Tristram_2022}. Several observational challenges need to be addressed in order to achieve tighter constraints: high sensitivity required by the faintness of the primordial B-mode signal and a high level of systematics control in order not to draw the faint polarized signal into instrumental polarization. Finally, the most significant challenge is the presence of polarized foregrounds from the Galaxy (dust, synchrotron, and potentially emission lines) or from gravitational lensing of the E-modes into B-modes. In particular, removing Galactic foregrounds requires a precise knowledge of their frequency spectra, which appears more complex than anticipated~\citep{Pelgrims_2021, regnier2023identifying}. This article shows how the latter challenge can be addressed in a specific manner by the QUBIC instrument.
    
    \setlength{\arrayrulewidth}{0.2mm}
    \setlength{\tabcolsep}{24pt}
    \renewcommand{\arraystretch}{1.2}
    \begin{table}[h!]
        \renewcommand{\arraystretch}{1.2}
        \begin{center}
            \caption{\label{tab: main_parameters}Main parameters of QUBIC Full Instrument (FI).}
            \begin{tabular}{p{5cm} m{1cm} m{1cm}}
                \hline
                \hline
                {\bf Parameters} & {\bf Channel~A} & {\bf Channel~B}\\
                \hline
                Frequency channels [GHz] \dotfill & 150 & 220 \\
                Bandwidth [GHz] \dotfill & 37.5 & 55 \\
                Effective FWHM [$^{\circ}$] \dotfill & 0.39 & 0.27 \\
                Number of detectors & $992$ & $992$ \\
            \end{tabular}
        \end{center}
    \end{table}
    
    The Q\&U Bolometric Interferometer for Cosmology (QUBIC)\footnote{\tt http://qubic.org.ar} is an instrument dedicated to measuring the CMB B-modes. It uses a novel technology called bolometric interferometry (BI), which combines the high sensitivity of cryogenic detectors with the high level of control of instrumental systematics provided by interferometry (see ~\cite{2020.QUBIC.PAPER1, 2020.QUBIC.PAPER2,2020.QUBIC.PAPER3} while the instrument is described in \cite{2020.QUBIC.PAPER4, 2020.QUBIC.PAPER5, 2020.QUBIC.PAPER6, 2020.QUBIC.PAPER7, 2020.QUBIC.PAPER8} for a series of articles describing QUBIC instrumentation and scientific forecasts). Table~\ref{tab: main_parameters} shows the main characteristics of the QUBIC Full Instrument (FI). A major feature brought by bolometric interferometry (BI) is the ability to perform spectral-imaging, which allows for splitting the physical bandwidth of the instrument, achieving up to 5 times higher spectral resolution~\citep{2020.QUBIC.PAPER2}. Note that this band-splitting occurs at the data analysis level and does not involve any hardware modification. This feature is crucial for controlling foreground contamination in B-mode maps by providing better measurements of the foreground spectral profile.
    
    This article focuses on the reconstruction of frequency maps for the QUBIC experiment using the specific feature of spectral-imaging. We will show how the two physical bands, respectively at 150 and 220 GHz can be split into sub-bands at the data analysis level in order to increase spectral resolution by projecting the measured Time-Ordered-Data (TOD) onto a number of sub-band maps. This requires accounting for the evolution of the synthesized beam of the instrument (the Point-Spread-Function) throughout the bandwidth, as well as correcting for bandpass integration of a sky composed by various astrophysical foregrounds.
    
    This article is organized as follows. The first part focuses on principles of the spectral-imaging, and how to add external data to regularize edge artifacts. The second part describes the results obtained from simulations. In the last section, we evaluate in a simplified manner the expected performance of this map-making technique on the primordial B-modes.
    
    In parallel and beyond the scope of this  work, we are also developing an algorithm that uses Bolometric Interferometry spectral capabilities directly performing map-making on astrophysical components, adjusting the components parameters at the same time. This is done in~\cite{cmm}, hereafter Component Map-Making (CMM).

    \section{Spectral Imaging with Bolometric Interferometry}

    After recalling some basic concepts about map-making in section~\ref{sec:classical_mapmaking}, we will detail in section~\ref{sec:synthesized_beam} how the frequency evolution of the QUBIC synthesized beam allows for retrieving frequency information within the physical bandwidth at the TOD level. In section~\ref{sec:spectral_imaging} we will explain how one can exploit this feature by projecting the wide-band TOD onto several sub-bands spanning the physical bandwidth of the instrument. We will describe a number of effects we have to account for in order to achieve an unbiased reconstruction of the sub-band maps.
    
    \subsection{Classical map-making and inverse-problem approach}
    \label{sec:classical_mapmaking}
    %short, just a few equations. Try to add references. Discuss inverse problem approach, PCG. Very general statements.
    CMB data are produced by observing the sky through the beam of the instrument according to a certain scanning strategy. The raw data is expressed in a very general way as:
    	\begin{equation}
            \vec{d} = \vec{H} \vec{s} + \vec{n},
            \label{eq:data}
        \end{equation}
    where $\vec{d}$ is the TOD, $\vec{s}$ is the true sky (I, Q, and U Stokes parameters maps) and $\vec{n}$ is the noise vector in TOD space. It includes two contributions: the intrinsic noise from the detectors and the noise from the incoming radiation (photon, atmospheric, environmental noise). The noise covariance matrix is $\vec{N}=\left<\vec{n} \cdot \vec{n}^T\right>$.
    The operator $\vec{H}$ is the acquisition matrix that describes the details of how the sky is observed by the instrument. It includes the pointing of the instrument, the convolution by the Point-Spread-Function (PSF), as well as a possible polarization modulation (from a Half-Wave-Plate for instance) and other instrumental effects such as the detector time constants.
    
    In the case of a classical imager, the instrumental beam is made of a single peak (the Airy disk for a circular primary mirror), so that, if $\vec{s}$ is the sky convolved at the angular resolution of the PSF, then at each time sample (corresponding to a given pointing of the telescope), a single pixel of the convolved sky is seen by any instrument detector. This results in $\vec{H}$ being a very sparse operator, with each line containing only zeros with a single one corresponding to the observed pixel. Due to the peculiar shape of the synthesized beam, this 1-to-1 correspondence between the sky and the samples does not apply for a Bolometric Interferometer such as QUBIC where the $\vec{H}$ operator is less sparse, as will be described below.
    
    If the noise is Gaussian with covariance matrix $\vec{N}$, one can analytically express the maximum likelihood estimate for the sky maps $\vec{\hat{s}}$ as:
        \begin{equation}
    	   \vec{\hat{s}} = \left( \vec{H}^{T} \vec{N}^{-1} \vec{H} \right)^{-1} \vec{H}^{T} \vec{N}^{-1} \vec{d}
            \label{eq:solution}
        \end{equation}
        
    However, even accounting for the sparsity of $\vec{H}$, it is difficult in most cases to calculate explicitly the solution given in Eq.~\ref{eq:solution} because of the size of the matrices that need to be inverted. Furthermore, it can be advantageous to refine significantly the operator $\vec{H}$ in order to account for a more detailed description of the actual instrument, including, for example, instrumental complexity and imperfections through Jones matrices \citep{O_Dea_2007}. In such a case one has to solve for $\vec{\hat{s}}$ without an explicit inversion. This can be performed using an inverse-problem approach where one uses a highly refined version of $\vec{H}$ that allows simulating the instrument acquisition in great detail from a simulated sky map $\vec{\Tilde{s}}$ (the $\tilde{~}$ will denote ``simulated data'' from now on):
    \begin{equation}
        \vec{\Tilde{d}} = \vec{H} \vec{\Tilde{s}},
    \end{equation}
    and then iterate the simulated sky map in order to minimize a cost function based on the difference between the simulated TOD and the measured data. The optimal solution is found minimizing:
    \begin{equation}
    \begin{split}
        \chi^2 & = \left( \vec{d} - \vec{\Tilde{d}}\right)^T \vec{N}^{-1} \left( \vec{d} - \vec{\Tilde{d}} \right) \\
        & = \left( \vec{d} - \vec{H} \vec{\Tilde{s}}\right)^T \vec{N}^{-1} \left( \vec{d} - \vec{H}\vec{\Tilde{s}} \right),
        \label{eq:cost_function}
    \end{split}
    \end{equation}
    where the multiplication by $\vec{N}^{-1}$ can be efficiently performed in Fourier space under the assumption of a stationary noise. One can use a preconditioned conjugate-gradient (PCG) algorithm in order to reach the actual solution of Eq.~\ref{eq:solution} through this inverse method. Even though this method does not involve inverting large matrices, it still requires significant computer resources for the application of $H$ to simulated data is performed at each step of the PCG and is generally performed on supercomputers.
    
    As mentioned above, in the case of a BI such as QUBIC, the complexity of the synthesized beam is such that the inverse problem approach is clearly advantageous and will be used throughout this article.

    \subsection{BI synthesized beam}
    \label{sec:synthesized_beam}
    
    In a BI such as QUBIC, the sky is not observed directly but instead through an array of apertures, like in a Fizeau interferometer \citep{2020.QUBIC.PAPER1}. The combination of the interference fringes formed by all pairs of apertures forms an image in the focal plane that corresponds to the ``dirty image'' of an interferometer. Each point in the focal plane observes the sky convolved by a PSF that depends on the shape of the array of apertures, their beams, and the location in the focal plane (because of changes in the phases of the interferences). 
    \begin{figure}[ht]
    	\includegraphics[scale=0.7]{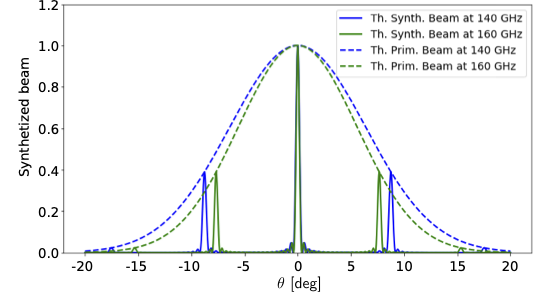}
    	\includegraphics[scale=1]{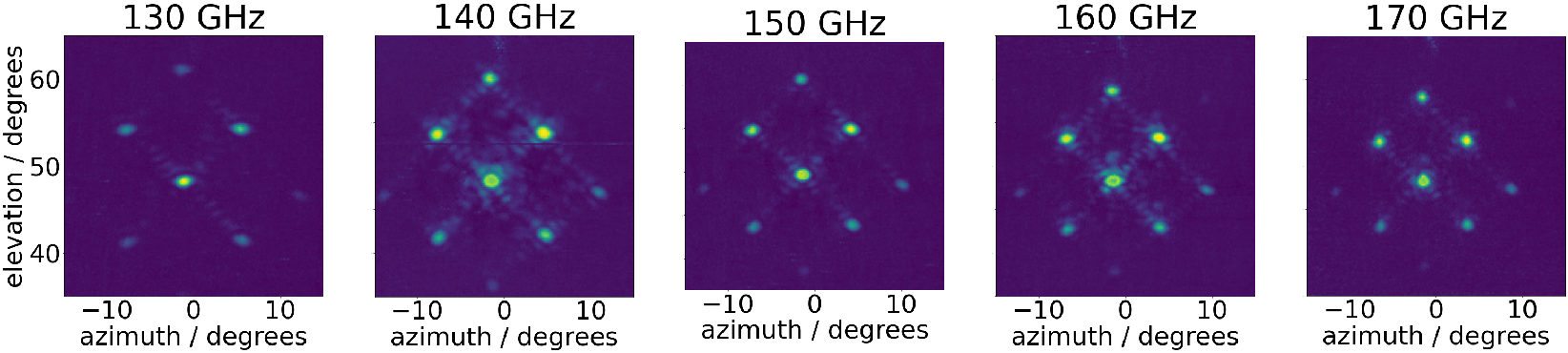}
    	\caption{\textit{Top panel}: Theoretical synthesized beam for a detector at the center of the QUBIC focal plane along with with its primary beam. $\theta$ is the angle between the detector axis and the observed direction. One can clearly see the frequency-dependent position of the secondary peaks. \textit{Bottom panel}: synthesized beam for one detector measured at various frequencies, from \cite{2020.QUBIC.PAPER3}. }
    	\label{fig:Theo_SB}
    \end{figure}
    A detailed description of the resulting PSF, called synthesized beam, is given in~\cite{2020.QUBIC.PAPER2}. The QUBIC synthesized beam is multiply-peaked, as shown in Fig.~\ref{fig:Theo_SB} where a cut at 45$^\circ$ through the theoretical synthesized beam is shown on the top and the frequency evolution of the location of the peaks is shown at the bottom. Each of the peaks is well-approximated by a Gaussian~\citep{stolp2016} with the same width, given by the ratio of the wavelength to the largest distance between two horns in the interferometer horn array. The angular distance between two peaks is the ratio between the wavelength and the distance separating two neighboring horns.
    
    With such a multiply-peaked PSF, it is clear that the usual map-making assumption of a line of zeros with a single 1 for the acquisition matrix $\vec{H}$ is not satisfied and so a specific algorithm needs to be developed. Our map-making technique uses the inverse-model approach described in section~\ref{sec:classical_mapmaking} with an operator $\vec{H}$ that incorporates the instrument's pointing and the actual direction and amplitude on the sky observed by each peak in the synthesized beam for each detector. $\vec{H}$ also accounts for the Half Wave-Plate (HWP) modulation on the sky signal and the optics of our BI.
    
\begin{figure*}[ht]
    	\includegraphics[width=\hsize]{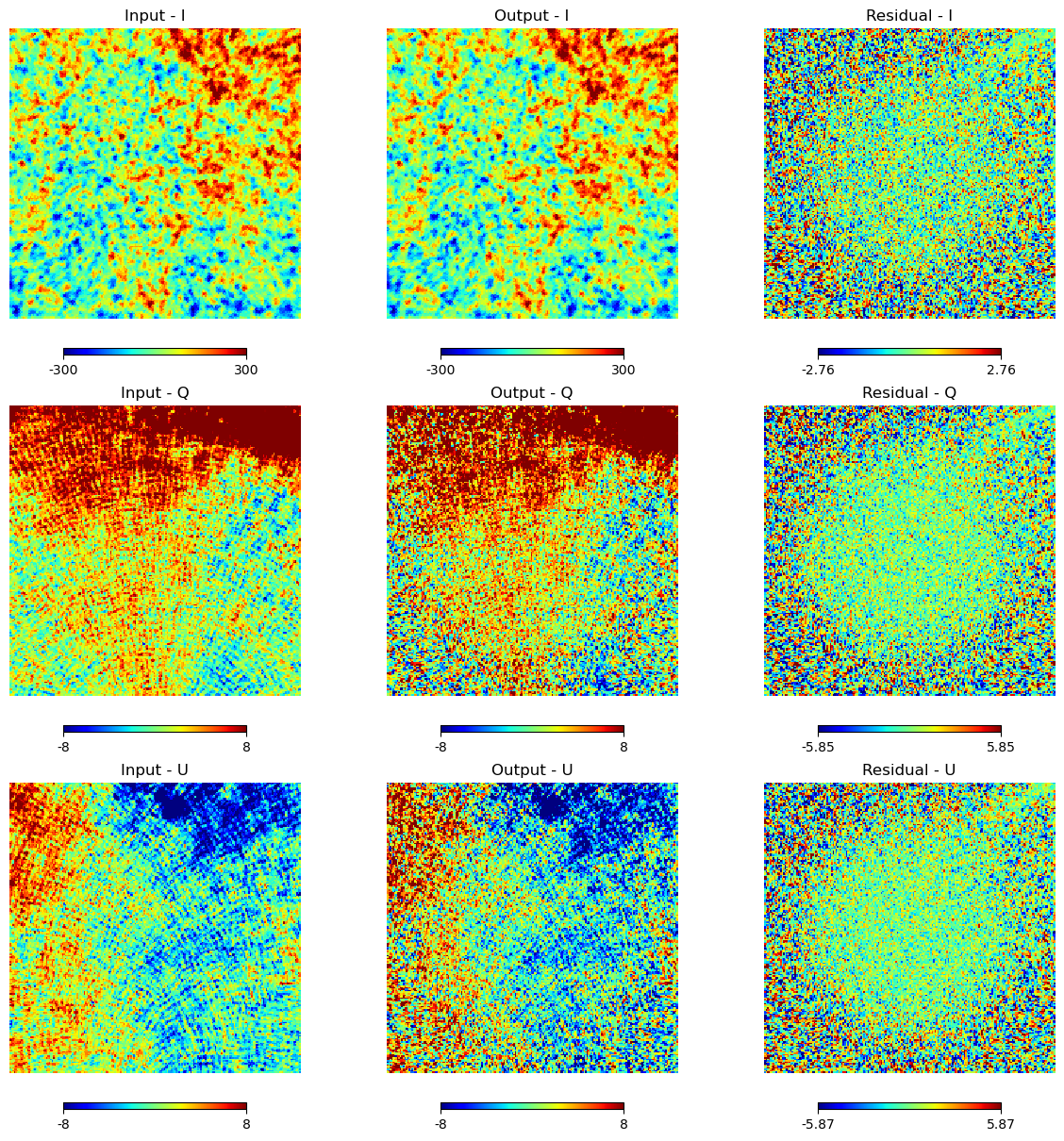}
    	\caption{Monochromatic case - From the top to the bottom: Stokes parameters I, Q, and U; from the left to the right: input, reconstructed maps, and the difference between them respectively, in $\si{\mu\K}_\text{CMB}$.}
    	\label{fig:res_mono}
    \end{figure*}

    For the sake of illustrating map-making with the QUBIC synthesized beam in a simple case, we show in Fig.~\ref{fig:res_mono} the reconstructed I, Q, and U maps in the case of a purely monochromatic synthesized beam at 150\,GHz, assuming 3 years of cumulative observations. This configuration define the ``nominal noise'', used in the rest of this work. Thermal dust emission is assumed here to be characterized by a Modified Black Body (MBB) law with $\beta_d = 1.54$. The noise has been extrapolated from the Technical Demonstrator (TD) measured noise level~\citep{2020.QUBIC.PAPER4}. We observe an unbiased reconstruction of the three Stokes parameters. 
    
    In a more realistic case, the synthesized beam evolves with frequency (as shown in the bottom panel of Fig.~\ref{fig:Theo_SB}). Accounting for this feature in the map-making is the basis of spectral-imaging map-making described in the next section. 
    
    \subsection{Spectral-Imaging Map-making}
    \subsubsection{General principle}
    \label{sec:spectral_imaging}
    In this section, we will neglect the effect of angular resolution, which will be incorporated in section~\ref{sect:angres}.
    The basic idea of spectral-imaging map-making is that, thanks to the change in the locations of the synthesized beam peaks as a function of frequency (as shown in Fig.~\ref{fig:Theo_SB}), a bolometric interferometer provides spectral information in the TOD~\citep{2020.QUBIC.PAPER2}. Because bolometers are only sensitive to the energy deposited by incoming photons, explicitly accounting for the variation of the synthesized beam within the filter bandwidth leads to replacing Eq.~\ref{eq:data} with
    
        \begin{equation}
            \vec{d} = \int_{\nu_{min}}^{\nu_{max}} \mathcal{H}_{\nu} \vec{s}_{\nu} \text{ d}\nu +\vec{n},
            \label{eq:data_int}
        \end{equation}
    
    \noindent where $\nu_{min}$ and $\nu_{max}$ are the physical bandwidth boundaries (we assume a perfectly flat band-pass for simplicity but a more realistic response of the band-pass filter can easily be accounted for), $\vec{s}_{\nu}$ is the real sky emitting at a given frequency $\nu$, and $\mathcal{H}_\nu$ is the monochromatic spectral acquisition operator at frequency $\nu$ that bears the information of the frequency evolution of the synthesized beam and the possible inclusion of systematics effects.
    
    We will define two important parameters, $N_{\text{rec}}$ and $f_{\text{sub}}$, respectively the number of reconstructed sub-band maps and the number of integration sub-bands within each reconstructed sub-band.
    The spectral imaging consists in solving for $N_\text{rec}$ sub-band maps $\vec{\tilde{s}}_{\nu_{i}}$ (the reconstructed maps) given a TOD $\vec{\tilde{d}}$ written as the sum of the energy deposited in each reconstructed sub-band:
    
        \begin{equation}
            \vec{\tilde{d}} = \sum_{i = 1}^{N_{\text{rec}}} \vec{H}_{\nu_i} \vec{\tilde{s}}_{\nu_i},
            \label{eq:data_sum}
        \end{equation}
    
    \noindent by minimizing the same cost function in TOD space as in Eq.~\ref{eq:cost_function}.
    The sub-band acquisition operator $\vec{H}_{\nu_i}$ is defined as the sum of the integration sub-acquisitions within the reconstruction sub-band $i$:
    
    \begin{equation}\label{defHsubband}
    \vec{H}_{\nu_i} = \sum_{j = 1}^{f_{\text{sub}}} \Delta\nu_{ij} \mathcal{H}_{\nu_{ij}},
    \end{equation}
    
    \noindent where $\Delta\nu_{ij}$ is the width of the discrete frequency intervals used in the model so that the polychromatic acquisition is written as
        \begin{equation}
            \vec{\tilde{d}} = \sum_{i = 1}^{N_{\text{rec}}} \left( \sum_{j = 1}^{f_{\text{sub}}}
            \Delta \nu_{ij} \mathcal{H}_{\nu_{ij}} \right) \cdot \vec{\tilde{s}}_{\nu_i}.
            \label{eq:spectral_im_model}
        \end{equation}
    
    The number of reconstructed sub-bands $N_\text{rec}$ is, of course, limited by the spectral resolution allowed by the bolometric interferometer, which is given by $\frac{\Delta\nu}{\nu} \simeq \frac{1}{P}$ where $P$ is the number of apertures in one dimension of the interferometer's pupil array \citep[see Eq. 3.1 from][]{2020.QUBIC.PAPER2}. QUBIC specific case should be able to reconstruct up to $N_\text{rec} \leq 5$. In order to smoothly describe the integral from Eq.~\ref{eq:data_int} with a sum as in Eq.~\ref{eq:data_sum}, one needs to use a larger number of integration sub-bands. Each of the integration sub-acquisition within a reconstructed sub-band will, however act on the reconstructed map $\vec{\tilde{s}}_{\nu_i}$, neglecting, for now, the sky variation within the reconstructed band (corrected for in section~\ref{sec:bandpass}).
    
    \subsubsection{Angular resolution}
    \label{sect:angres}
    The monochromatic synthesized beam of QUBIC has multiple peaks and their locations are the basis for spectral imaging. The angular resolution is set by the width of each of the peaks, well approximated by a Gaussian with $\sigma_j=\lambda_j/{\Delta x}$, where $\lambda_j$ is the wavelength at which the synthesized beam is considered, and $\Delta x$ is the distance separating two pupils in the interferometer~~\citep{stolp2016, 2020.QUBIC.PAPER2}. Since the angular resolution is different for each sub-frequency we are considering, convolutions to the actual angular resolution need to be incorporated into the acquisition model. If we define the sky at frequency $\nu_j$ and with ``infinite'' resolution as $\vec{s}_{\nu_j}^\infty$, the model for the measured data becomes:
        \begin{equation}
            \vec{d} = \int_{\nu_{min}}^{\nu_{max}} \mathcal{H}_{\nu} C_\nu \vec{s}_{\nu}^\infty \text{ } d\nu +\vec{n},
            \label{eq:data_angres}
        \end{equation}
    where $C_\nu$ is the convolution operator (Gaussian kernel) that transforms the infinite resolution sky to the resolution at frequency $\nu$.
    
    Reconstructing maps at infinite resolution is impractical. Instead, the resolution for each reconstructed map is chosen as the optimal resolution, corresponding to the highest frequency within the reconstructed sub-band, denoted as $\sigma_{\nu_i}$. Furthermore, the integration operators within the reconstructed sub-band $j$ need to act on a map with matched resolution $\sigma_{\nu_{ij}}$. To achieve corresponding resolutions, degradation using a Gaussian kernel $\mathcal{C}_{K_{ij}}$ is incorporated, with the kernel width given by:
    \begin{equation}
        \sigma_{K_{ij}} = \sqrt{\sigma_{\nu_i}^2 - \sigma_{\nu_{ij}}^2}.
    \end{equation}
    The final model for the simulated data constructed during the PCG iterations therefore becomes:
        \begin{equation}
            \vec{\tilde{d}} = \sum_{i = 1}^{N_{\text{rec}}} \left( \sum_{j = 1}^{f_{\text{sub}}}
            \Delta\nu_{ij} \mathcal{H}_{\nu_{ij}} \mathcal{C}_{K_{ij}} \right) \cdot \vec{\tilde{s}}_{\nu_i},
            \label{eq:data_conv_rec}
        \end{equation}
    where the reconstructed sky $\vec{\tilde{s}}_{\nu_i}$ is now at angular resolution $\sigma_{\nu_i}$.
    
    Accounting for these resolutions considerably slows down the PCG iterations, as convolutions need to be performed at each step. However, we have parallelized this process so that it is doable in practice.
    
    In appendix~\ref{sec: appendixB}, we explore the possibility of not doing those convolutions during reconstruction, in order to speed up the calculations. The PCG will naturally fit each reconstructed map of the $N_\text{rec}$ sub-bands to an average resolution over the $f_\text{sub}$ frequencies of each sub-band. Analytical approximation of the resolution allows for the cosmological analysis, as described in section~\ref{multicomps}. This calculation also gives a precise definition of the average frequency $\nu_i$.
    
    \subsubsection{Bandpass Binning Bias}
    \label{sec:bandpass}
    
    The other major effect that is very important to understand and take into account is the so-called bandpass binning bias. The CMB itself, expressed in temperature units, does not vary within the bandwidth, but the measured (or ``observed'') sky contains foregrounds that significantly vary within the band.
    
    In our case, the mismatch between the measured data and our reconstruction model arises from neglecting the variation of the emission from the sky within each reconstructed sub-band in Eq.~\ref{eq:spectral_im_model}, where a unique sky $\vec{\tilde{s}}_{\nu_i}$ is used for all the integration sub-frequencies of this reconstructed sub-band (index $i$). Although small, this effect produces non-negligible residuals on the reconstructed maps. The smaller the number of reconstructed maps, the higher the impact, as the constant map approximation gets worse and worse when increasing the bandwidth of the reconstructed sub-bands.
    
    In order to minimize the bandpass binning bias effect, we have used a model dependent method inspired by~\cite{svalheim2022beyondplanck}. We used a simulated sky model from PySM\footnote{{\tt https://pysm3.readthedocs.io/en/latest/}}~(\cite{Zonca_2021}, \cite{Thorne_2017}), containing only foregrounds (as the CMB does not induce mismatch), and calculate from this model the difference for each reconstructed sub-band between the average map in the sub-band (index $i$) and the actual integration sub-maps (index $j$):
    \begin{equation}
        \vec{\delta}_{ij} = \left\langle \vec{s}_{\nu_{ij}}^{\mathrm{PySM}}\right\rangle_{i} - \vec{s}_{\nu_{ij}}^{\mathrm{PySM}}.
    \end{equation}
    
    We can now convert this map mismatch into time domain by applying our operators in order to construct a time-domain correction $\vec{\delta}$:
    \begin{equation}
    \vec{\delta} = \sum_{i = 1}^{N_{\text{rec}}} \sum_{j = 1}^{f_{\text{sub}}} \Delta\nu_{ij}\mathcal{H}_{\nu_{ij}} \mathcal{C}_{K_{ij}} \vec{\delta}_{ij},
    \end{equation}
    which is subtracted from our time-ordered data before being input into the map-making algorithm:
    \begin{equation}
        \vec{d}_\mathrm{corr} = \vec{d} - \vec{\delta},
        \label{eq:bandpass_mismatch}
    \end{equation}
    resulting in a negligible bias on the reconstructed maps in our simulations based on PySM sky.
    
    This method relies on using a sky model for the bandpass correction and could, therefore, result in systematics if the sky model is incorrect. Especially, wrongly correcting for this bias can induce errors on multi-frequency analysis, describe later in this work. One will need to pay extra care in this part of the algorithm when reconstructing real-sky data, as well as use the state-of-the-art sky model available at the time of analyzing real data and possibly iterate between map-making and improved sky model in order to minimize the mismatch. It is also possible to extract spectral information from the results of components map-making~\citep{cmm} to evaluate the observed spectral energy distribution (SED) and correct frequency map-making through an iterative scheme.
    
    \subsection{Adding external data}\label{section:ext}
    
    The data model presented so far only includes QUBIC data. This leads to undesired boundary effects. When reconstructing a pixel located near the edges of the observed patch, information is needed from pixels situated in the location of all peaks forming the synthesized beam, a few degrees apart from each other. Some of these peaks are located deeper within the observed patch ($15^{\circ}$ radius circle centered on $\left[0, -57^{\circ}\right]$) and, therefore, provide relevant information. Other peaks fall even farther form the center of the patch, an area where QUBIC does not have enough information. This results in significant reconstruction errors around the edges of the observed patch shown in Fig.~\ref{fig:profile_broadband} (blue curves).
    
    Solving this issue is possible by optimally combining our data with external datasets (namely Planck frequency maps in intensity and polarization)  in order to incorporate knowledge about the pixels, poorly observed by QUBIC, but entering the reconstruction of pixels within our patch. In order to do so, we build a generic model to incorporate external data such as Planck:
    
    \begin{equation}
        H_\text{Tot} = \begin{pmatrix}
            \vec{H}^{\text{QUBIC}}_{\nu_1} & \vec{H}^{\text{QUBIC}}_{\nu_2} & \cdots & \vec{H}^{\text{QUBIC}}_{\nu_{N_\text{rec}}} \\
            \vec{H}^{\text{Ext}}_{\nu_1} & 0 & 0 & 0 \\
            0 & \vec{H}^{\text{Ext}}_{\nu_2} & 0 & 0 \\
            0 & 0 & \ddots & \vdots \\
            0 & 0 & \cdots & \vec{H}^{\text{Ext}}_{\nu_{N_\text{rec}}} 
            \end{pmatrix}
            = \begin{pmatrix}
                \vec{H}^{\text{QUBIC}} \\
                \vec{H}^{\text{Ext}}
            \end{pmatrix},
            \label{eq:H_poly}
    \end{equation}
    where $\vec{H}^{\text{QUBIC}}_{\nu_i}$ is defined in Eq.~\ref{defHsubband} denotes the integral of the QUBIC operator within the reconstructed sub-band $i$. $\vec{H}^{\text{Ext}}_{\nu_i}$ is a much simpler operator that only reads a sky map constructed from the external data to cover the reconstructed sub-band $i$. In practice, we use Planck data and build each of these sub-band maps from the component-separated maps from Planck~\citep{PlanckIV2020} with the components mixing matrix corresponding to this band. We do so in order to be able to deal with any sub-band splitting that, in general, will not match the actual frequency bands of Planck.
    
    The data will therefore be a larger vector containing first the QUBIC data as before and then the various external TOD constructed as explained above:
    \begin{equation}
        \vec{d} = \begin{bmatrix}
            \vec{d}_\text{QUBIC}\\
            \vec{d}_\text{Ext}
        \end{bmatrix}.
    \end{equation}
    
    With such a model, one can benefit from the spectral-imaging capabilities of QUBIC without any restriction in the choice of the reconstructed frequency sub-bands, while benefiting from the external data for the pixels for which QUBIC has only poor information.
    
    We ensure the near-optimality of our model by introducing weights for the external data computed from the Planck publications. Those weights are artificially set to zero inside the QUBIC patch in the noise covariance matrix of Planck data, so that we minimize the possible systematics from the external dataset, restricting its use to lifting degeneracies for the pixels near the edges of the patch that require information from outside the patch.
    
    The final cost function we minimize in our forward modeling reconstruction is as before, where $\vec{\tilde{s}}$ are the unknown sub-bands maps:
    \begin{equation}
    \begin{split}
        \chi^2_\text{Tot} & = \left( \vec{d} - \vec{\tilde{d}}\right)^T \vec{N}^{-1} \left( \vec{d} - \vec{\tilde{d}}\right)\\
        & = \left( \vec{d} - \vec{H}_\text{Tot} \vec{\tilde{s}}\right)^T \vec{N}^{-1} \left( \vec{d} - \vec{H}_\text{Tot} \vec{\tilde{s}}\right).
    \end{split}
    \end{equation}
    
    The cost function explicitly accounts for both QUBIC and the External data contributions, whose noises are not correlated:
    \begin{align*}
        \chi^2_\text{Tot} &= \begin{bmatrix}
                        \vec{d}_\text{QUBIC} - \vec{\tilde{d}}_\text{QUBIC} \\
                        \vec{d}_\text{Ext} - \vec{\tilde{d}}_\text{Ext} \\
                    \end{bmatrix}^T
                    \begin{bmatrix}
                        \vec{N}^{-1}_\text{QUBIC} & 0 \\
                        0 &  \vec{N}^{-1}_\text{Ext} \\
                    \end{bmatrix} 
                    \begin{bmatrix}
                        \vec{d}_\text{QUBIC} - \vec{\tilde{d}}_\text{QUBIC} \\
                        \vec{d}_\text{Ext} - \vec{\tilde{d}}_\text{Ext} \\
                    \end{bmatrix} \\
                 &= \left( \vec{d}_\text{QUBIC} - \vec{H}^{\text{QUBIC}} \vec{\tilde{s}}\right)^T \vec{N}^{-1}_\text{QUBIC} \left( \vec{d}_\text{QUBIC} - \vec{H}^{\text{QUBIC}}\vec{\tilde{s}} \right) \\
                 &~~~~~+
                 \left( \vec{d}_\text{Ext} - \vec{H}^{\text{Ext}} \vec{\tilde{s}}\right)^T \vec{N}^{-1}_\text{Ext} \left( \vec{d}_\text{Ext} - \vec{H}^{\text{Ext}}\vec{\tilde{s}} \right) \\
                 &= \chi^2_\text{QUBIC} + \chi^2_\text{Ext}.
    \end{align*}

    The above cost function is computed for the full-sky as we add Planck data but QUBIC is only added on its patch. The regions outside the QUBIC patch is only used to regularize the edge effects, and will be mask for the cosmological inference (see section~\ref{sec: analysis principle}). We want to emphasize that the benefits of using external data come only from observing the same portion of the sky with different instruments at different frequencies.
    
    \section{Simulations and results}

    In this section, we first present the results obtained during end-to-end simulations in the case of a broadband acquisition, meaning that we do not attempt to perform spectral-imaging but rather a single map integrated throughout the physical bandwidth of the instrument, accounting for the multiple peaks of the synthesized beam and its evolution as a function of frequency, as well as including external data in order to solve for the edge effects mentioned above. The setup for these simulations relies on simplistic assumptions as random pointing, relatively simple noise, no systematics. Secondly, we present results based on the same simulations, but with a polychromatic reconstruction, meaning that we perform spectral-imaging, projecting the TOD onto a number of sub-bands within the physical bandwidth of the instrument.
    Finally, we will focus on characterizing the noise structure in our reconstructed maps and the expected performance of the instrument.
    
    \subsection{Broadband acquisition}
    
    The full sensitivity of the instrument can be seen with the broadband acquisition which gathers all the photons collected by the instrument and projects them onto a single, broadband map. During deconvolution by multiple peaks, the position of each peak is taken into account for each frequency. The signal simulated in this section is composed of a CMB primordial fluctuation signal and a thermal dust signal. Following  \cite{2020.QUBIC.PAPER1} (Fig.~10), we do not include synchrotron emission as it is negligible at our frequencies.
    
    \begin{figure*}
        \centering
    	\includegraphics[scale = 0.4]{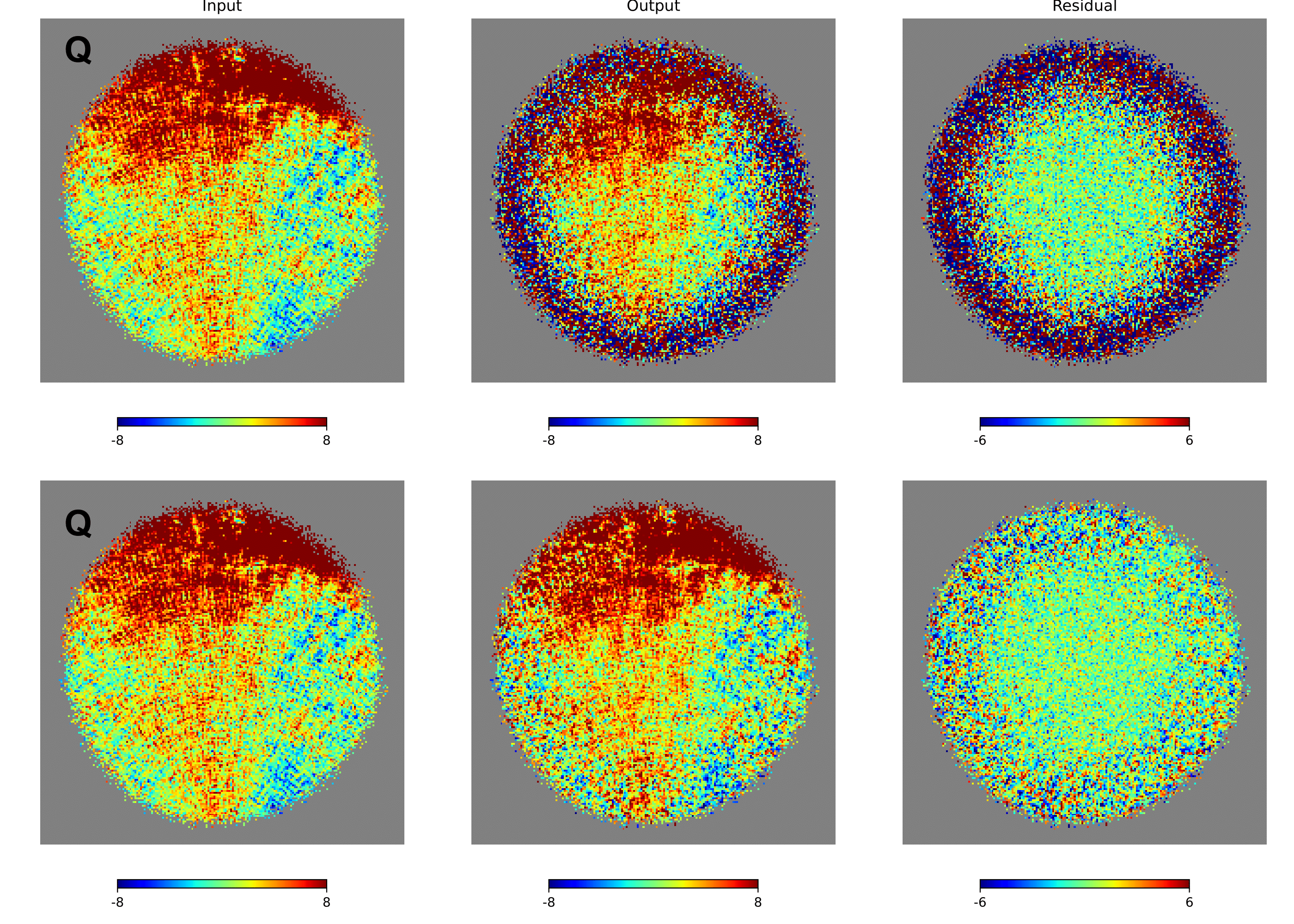}
    	\caption{Broadband case - First row is the result of the QUBIC acquisition only with the edges effect. The second row is the merging of the QUBIC and Planck acquisitions. From the left to the right is the input, output and the difference between them respectively for the Q Stokes parameter, in $\si{\mu\K}_\text{CMB}$. This simulation integrates the 150 GHz band from 131 GHz to 169 GHz (see Table~\ref{tab: main_parameters}).}
    	\label{fig:maps_broadband}
    \end{figure*}
    
    The integration over the whole bandpass is done by a sum of sub-operators each describing the synthesized-beam at a given frequency. We integrate using the trapezoidal method, fixing the reconstructed angular resolution according to the frequency. After proceeding to the correction of the TOD following the procedure described in section~\ref{sec:bandpass}, we minimize our $\chi^2$ function. The resulting maps are presented in Fig.~\ref{fig:maps_broadband}. The first line shows the reconstruction using QUBIC acquisition alone with, from left to right, the input Q map, the reconstructed Q map, and the residuals respectively. 
    As expected, significant edge effects can be observed in this case, although similar to noise, they are caused by the multiple-peak deconvolution involving peaks pointing far from the center of QUBIC's patch, where the low amount of information causes large degeneracies.
    
    The second line shows the reconstruction of the same TOD when Planck data are accounted for as explained in section~\ref{section:ext} in order to regularize the edge effects. The degeneracies disappear showing an improved noise near the edges of the map.  Note that we force the Planck weight, given by the inverse noise covariance matrix, to zero in the inner part of the QUBIC patch. The zone where we only keep QUBIC data is typically where the normalized coverage is above 20\%, this is where QUBIC is able to perform spectral-imaging. Convergence is achieved using a PCG method, updating the pixels at each iteration of the PCG towards the solution that minimizes our cost function.
    
    %\begin{figure}
    %    \centering
    %	\includegraphics[scale = 0.5]{figure/profile_broadband.pdf}
    %	\caption{Broadband case - Profile of the residual RMS for Q (\textit{top}) and U \textit{(bottom)} Stokes parameter. The blue line shows the RMS using QUBIC only and the red line shows the RMS when using  Planck data to regularize the edge effects.}
    %	\label{fig:profile_broadband}
    %\end{figure}
    
    \subsection{Polychromatic reconstruction}
    
    The pointing matrix which is our reconstruction model shown by the equation~\ref{eq:H_poly}, is numerically complicated to compute due to the large memory requirement. The computations done in this article take about ten hours per realization, using 4 cores with 4 CPUs each. In each core, the focal planes are divided using MPI~\citep{DALCIN20051108} into several subsets of detectors that are jointly analyzed. For regularization, we use reconstructed Planck astrophysical components recombined according to the best-fit mixing matrix at the relevant frequency.
     The band integration bias is taken into account according to the equation~\ref{eq:bandpass_mismatch}. The use of Planck data speeds up our total convergence by reducing the number of iterations until a threshold is reached indicating fairly accurate convergence. 
    
    Spectral-imaging is a flexible technique that allows one to increase the spectral resolution. This information is particularly important as the actual spectral complexity of astrophysical foregrounds is still unknown. In CMM, we explore the even more powerful possibility of performing component separation at the map-making stage, fully benefiting from the spectral-imaging capabilities of BI.
    
    \begin{figure}
        \centering
        %\hspace{-1cm}
    	\includegraphics[scale = 0.60]{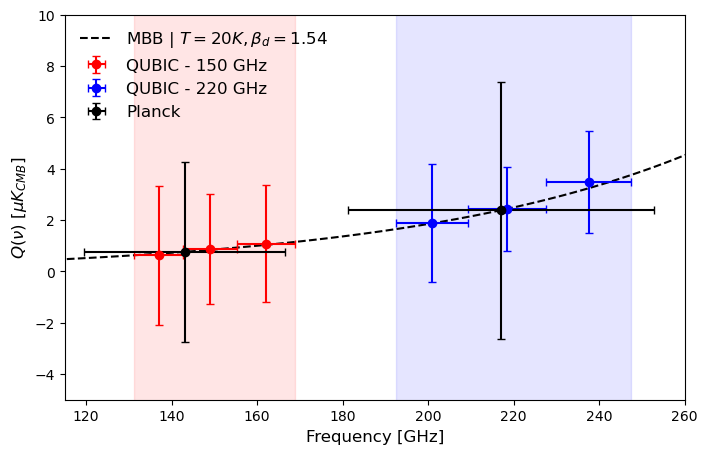}
    	\caption{Reconstructed spectral energy distribution (SED) of Q~Stokes parameter for $N_{\text{rec}} = 3$ for both physical bands. Dots are the value of a given pixel in the QUBIC patch and errorbars are the dispersion (RMS) over the whole QUBIC patch. Black dots denote Planck uncertainties on the QUBIC patch. The dashed-line is the MBB spectrum of the dust. The pixelization of the sky has been set to $N_{\text{side}} = 256$.}
    	\label{fig:sed_poly}
    \end{figure}
    
    Considering the same simulated TOD as in the previous section, it is possible to discretize the reconstruction no longer by reconstructing an ``averaged'' map in the band, but a set of maps whose recombination through the operator $\cal{H}$ corresponds to the TODs. We performed simulations considering only thermal dust emission which increases with respect to the frequency. Fig.~\ref{fig:sed_poly} shows the reconstructed SED for the Q Stokes parameters. To constrain the convergence of the 150 and 220~GHz bands of QUBIC, we use the 143 and 217~GHz bands of Planck. The black window shows the sensitivity of Planck data with one map within each physical band. The ability of QUBIC to do spectral-imaging permits the increase of spectral resolution (considering $N_{\text{rec}} = 3$ per physical bands) and achieves better constraints on astrophysical foregrounds, a major challenge for B-modes search, in particular in the case of  complex foregorunds foregrounds such as frequency decorrelated thermal dust~\citep{regnier2023identifying}.

    \subsection{Noise characterization}
    \label{noise_characterization}
    
    We study the noise reconstruction through our pipeline by considering a sky with no signal, just instrumental noise. We take into account the detector noise with a Noise Equivalent Power $\text{NEP}_{\text{det}} = 4.7 \times 10^{-17} \text{ W}/\sqrt{\text{Hz}}$~\citep{2020.QUBIC.PAPER4} and the photon noise associated with each frequency, described by the $\text{NEP}_{\gamma}$ coming from all the sub-systems. The atmosphere is considered stable. This noise is reconstructed with different numbers of reconstruction sub-bands. 
    
\begin{figure*}
        \centering
        %\hspace{-1cm}
    	\includegraphics[scale = 0.55]{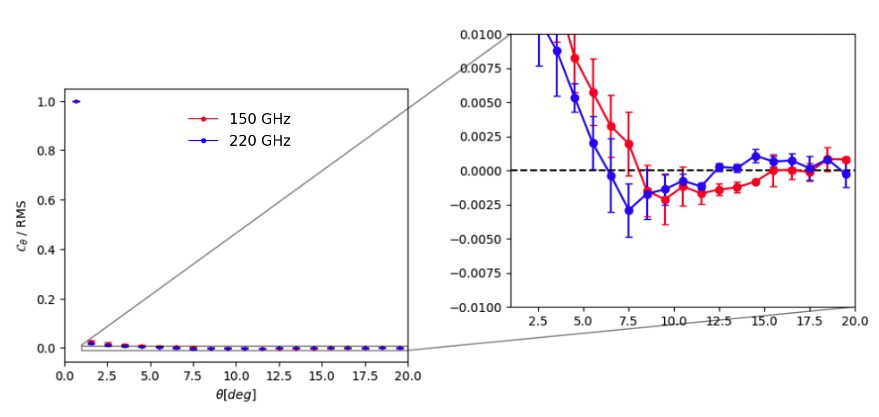}
    	\caption{2-point correlation function obtained from end-to-end simulations assuming $N_{\text{rec}} = 1$. Red and blue color correspond to the 150 GHz and 220 GHz bands respectively.}
    	\label{fig:ctheta}
    \end{figure*}

    In the first simple case, we consider $N_{\text{rec}} = 1$, which allows the reconstruction of one map per physical band, in the manner of a classical imager. We performed simulations with independent realizations at the noise level expected for QUBIC in order to study the properties of the noise in the reconstructed maps. We show the residual's  2pt-correlation function normalized by the RMS $\mathcal{C}(\theta = 0)$ in Fig.~\ref{fig:ctheta}. The correlation is almost zero as angular distance increases, indicating very small correlations in the noise structure and therefore an almost white noise. A residual pixel-pixel correlation remains at angles below $\sim$8.5 degrees at 150 GHz. This angular scale corresponds to the interpeak distance in the synthesized beam at 150 GHz distance. The residual correlation arises from the deconvolution of the multiple peaks performed by our map-making algorithm. A similar effect is observed at 220 GHz, at a slightly smaller angular scale because of the frequency scaling of the synthesized beam (see Fig.~\ref{fig:Theo_SB}). This almost-white structure of our noise may seem contradictory to what we found in~\cite{2020.QUBIC.PAPER1}. In fact, the strong correlations observed in that study were driven by the edge effect which we successfully regularized using external data in the present article.
    
\begin{figure*}
        \centering
        %\hspace{-1cm}
    	\includegraphics[scale = 0.35]{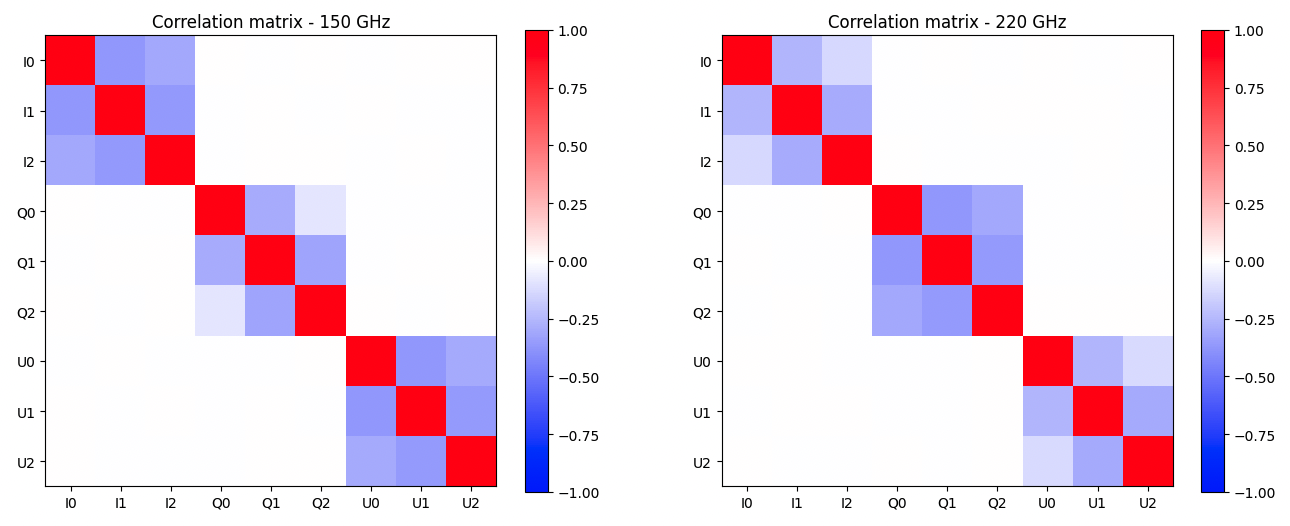}
    	\caption{Correlation matrix for noise reconstruction assuming $N_{\text{rec}} = 3$ per physical band at 150 GHz (left) and at 220 GHz (right). We note in the labels the Stokes component and the number of the reconstructed sub-band. Matrices are estimated by averaging over all the pixels seen by QUBIC.}
    	\label{fig:correlation}
    \end{figure*}

    We then produced similar simulations but by varying the number of reconstructed sub-bands ($N_{\text{rec}}>1$). Reconstructed maps within a physical band are statistically independent from those in the second physical band. 
    As the spectral information is extracted from the frequency-dependent description of the multiply peaked shape of the synthesized beam, we expect noise anti-correlation to appear between neighboring reconstructed sub-bands. The correlation matrices between the same pixel across the reconstructed sub-bands are shown in Fig.~\ref{fig:correlation} and were measured from our simulations. There is no visible correlation between the Stokes parameters but a certain level of anti-correlation between the sub-bands, as expected. Such correlations need to be accounted for in the subsequent analyses. The non-diagonal and anti-correlated nature of the band-band covariance matrices also implies that a pure measurement of the RMS in sub-band maps is not a good indicator of the actual noise in our set of reconstructed maps. Anti-correlations will enhance the RMS in each map, but in an anti-correlated manner, that can be accounted for.
    
    \section{Cosmological and foreground parameters analysis}
    \label{multicomps}
    
    The reconstruction described in the previous sections is very time-consuming but allows frequency maps to be reconstructed. The next step for any CMB experiment would be to apply a component separation method to extract a ``clean'' cosmological signal~\citep{Stompor}. However, this step would require accounting for the observed anti-correlations between the bands, which is possible but requires developments beyond the scope of this article. Another possibility, which will be the focus of the CMM article, is to perform component separation simultaneously with map-making, fully benefiting from the spectral imaging capabilities of BI~\citep{cmm}. For the sake of simplicity, we use here a method based on power spectra and cross-power spectra of the reconstructed maps, similar to what has been used in~\cite{Ade_2021}. A full Monte-Carlo analysis with forecasts for QUBIC and a number of sub-bands with various foreground models is on the way, and will be published separately. For now, this analysis allows us to assess the feasibility of our BI map reconstruction approach.
    
    \subsection{Principle}\label{sec: analysis principle}
    We use simulated QUBIC data at 150 and 220\, GHz to reconstruct maps, using Planck to regularize the edge effects. During the reconstruction, we ensure that Planck data are not used within the QUBIC patch to not double count the information in the likelihood analysis. Our analysis is done in angular power spectrum space, meaning that we have $N_{\text{rec}} \left( N_{\text{rec}} + 1\right) / 2$ spectra, all biased by the noise covariance of the corresponding maps. Eq.~\ref{eq:MCT_eq} describes the emission for each pair of frequencies with $\ell_0 = 80$ the reference multipole. 
    \begin{equation}
    \begin{split}
        \mathcal{D}_{\ell}^{\nu_1 \times \nu_2} ={} & r \times \mathcal{D}_{\ell}^{\text{tensor}} + A_{\text{lens}} \times \mathcal{D}_{\ell}^{\text{lensed}} \\
        & + A_d \Delta_d f_d^{\,\beta_d}(\nu_1) f_d^{\,\beta_d}(\nu_2) \left( \frac{\ell}{\ell_0} \right)^{\alpha_d} \\
        & + A_s \Delta_s f_s^{\,\beta_s}(\nu_1) f_s^{\,\beta_s}(\nu_2) \left( \frac{\ell}{\ell_0} \right)^{\alpha_s} \label{eq:MCT_eq} \\
        & + \varepsilon \sqrt{A_d A_s} \left( f_d^{\,\beta_d}(\nu_1) f_s^{\,\beta_s}(\nu_2) + f_s^{\,\beta_s}(\nu_1) f_d^{\,\beta_d}(\nu_2)\right) \left( \frac{\ell}{\ell_0} \right)^{\frac{\alpha_d + \alpha_s}{2}}
    \end{split}
    \end{equation}
    
    The model includes the CMB component by treating primordial perturbation and lensing residuals respectively with $r$ and $A_{\text{lens}}$. We describe the foreground emission as a power-law in \mbox{$\ell$-space}, the dust emission is defined by a Modified Black Body (MBB) in frequency space. The model includes the synchrotron contribution but given the frequencies and sensitivities we deal with, synchrotron is negligible~\citep{2020.QUBIC.PAPER1} and the corresponding parameters are fixed to zero. Further studies can be done in order to incorporate low-frequency data to constrain synchrotron emission. We can also consider the spatial correlation between dust and synchrotron patterns in the sky through a single parameter $\varepsilon$. The model can be summarized as follows:
    
    \begin{enumerate}
        \item $r$: Tensor-to-scalar ratio
        \item $A_{\text{lens}}$: Gravitational lensing residual
        \item $A_d$: Dust amplitude at 353 GHz
        \item $\alpha_d$: Dust spatial index
        \item $\beta_d$: Dust spectral index
        \item $\Delta_d$: Dust frequency decorrelation
        \item $A_s$: Synchrotron amplitude at 23 GHz
        \item $\alpha_s$: Synchrotron spatial index
        \item $\beta_s$: Synchrotron spectral index
        \item $\Delta_s$: Synchrotron frequency decorrelation
        \item $\varepsilon$: Spatial correlation between Dust and Synchrotron
    \end{enumerate}
    
    The MBB spectrum of the dust and the power law of the synchrotron are defined by:
    \begin{align}
        f_d^{\,\beta_d}(\nu) & = \frac{e^{\frac{h\nu}{kT_d}}-1}{e^{\frac{h\nu_0}{kT_d}}-1}\left(\frac{\nu}{\nu_0}\right)^{1+\beta_d} \cdot \frac{f_\text{CMB}(\nu_0)}{f_\text{CMB}(\nu)}\\
        f_s^{\,\beta_s}(\nu) & = \left(\frac{\nu}{\nu_0}\right)^{\beta_s} \cdot \frac{f_\text{CMB}(\nu_0)}{f_\text{CMB}(\nu)},
    \end{align}
    with $\nu_0$ a reference frequency, and the temperature of the dust usually set to $T_d=\SI{20}{\K}$. $f_\text{CMB}$ is a conversion factor to express the maps in units $\si{\mu\K}_\text{CMB}$:
    \begin{equation}
        f_\text{CMB}(\nu) = \frac{e^{\frac{h\nu}{k T_\text{CMB}}} \left(\frac{h\nu}{k T_\text{CMB}}\right)^2}{\left(e^{\frac{h\nu}{k T_\text{CMB}}}-1\right)^2}.
    \end{equation}
    
    We employ the Monte Carlo Markov Chain (MCMC)\footnote{{\tt https://emcee.readthedocs.io/en/stable/}} method for parameter estimation due to its capability to explore complex parameter spaces and handle correlations between parameters efficiently. MCMC allows sampling posterior distribution of model parameters, providing robust estimates and uncertainties. Spectra are binned using the NaMaster\footnote{{\tt https://github.com/LSSTDESC/NaMaster/tree/master}} package \citep{Alonso_2019} that includes purification of B-modes for the BB power spectrum, we used for our simulations a constant binning set to $\Delta \ell = 30$ on a multipole range from $\ell_{\text{min}} = 40$ to $\ell_{\text{max}} = 2 \times N_{\text{side}}$ with $N_{\text{side}} = 256$ using HEALPix\footnote{{\tt https://github.com/healpy/healpy}} map pixelization \citep{Gorski_2005}.
    
    \subsection{Likelihood analysis}
    
    To estimate noise-related uncertainties on power spectra, we create noise-only simulations using map making techniques for estimating the noise covariance matrix independently of any foreground effects that are systematics. To account for the sample variance of the CMB and foreground in the budget, we use the formula expressed in \cite{trendafilova2023impact} and developed in Appendix~\ref{sec: appendixA} as:
    \begin{equation}
        \text{Cov}_{\ell_1 \ell_2} \left( \mathcal{D}_{\ell}^{\nu_i \times \nu_j}, \mathcal{D}_{\ell}^{\nu_k \times \nu_l} \right) = \delta_{\ell_1 \ell_2} \frac{\mathcal{D}_{\ell_1}^{\nu_i \times \nu_k} \mathcal{D}_{\ell_1}^{\nu_j \times \nu_l} + \mathcal{D}_{\ell_1}^{\nu_i \times \nu_l} \mathcal{D}_{\ell_1}^{\nu_j \times \nu_k}}{\left( 2 \ell + 1\right) f_{\text{sky}} \Delta \ell},  \label{eq: samplevariance}
    \end{equation}
    where $\mathcal{D}_{\ell}^{\nu_j \times \nu_j}$ is the cross-spectra between corresponding maps at frequency $\nu_i$ and $\nu_j$, $f_{\text{sky}} = 0.015$ is the observed sky fraction, and $\Delta_{\ell}$ the binning of the spectrum. Note that this term is only added to the diagonal of the covariance matrix. 
    
    \begin{figure*}
        \centering
        \includegraphics[scale=0.5]{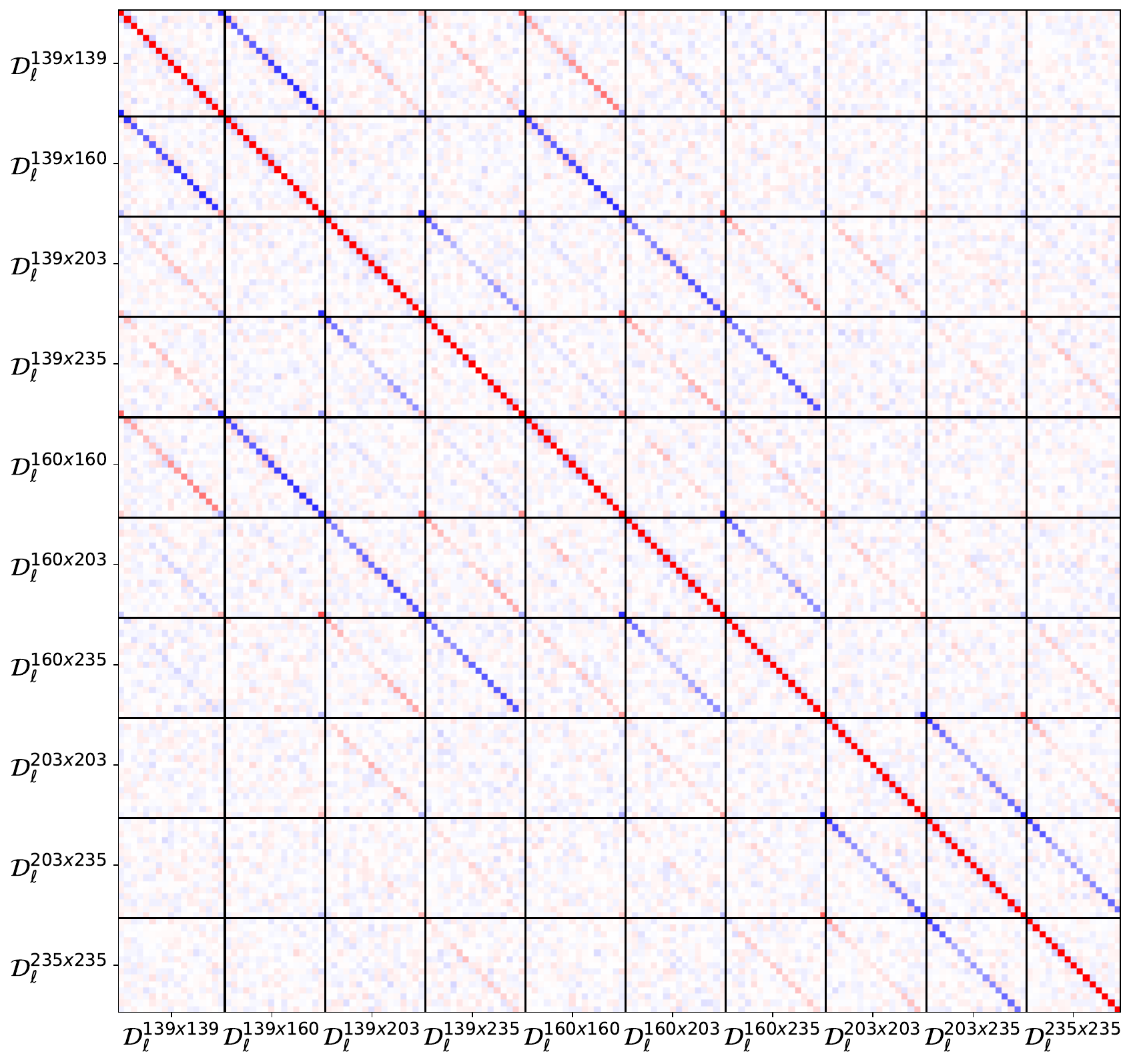}
        \caption{Correlation matrix for 10 pairs of cross-spectrum assuming 4 QUBIC's reconstructed maps (2 for each physical band). Positive and negative correlations are shown by red and blue colors respectively. The color scale goes from -1 (blue) to 1 (red).}
        \label{fig: corr_nrec4}
    \end{figure*}
    
\begin{figure}%[H]
        \centering
        \includegraphics[scale = 0.55]{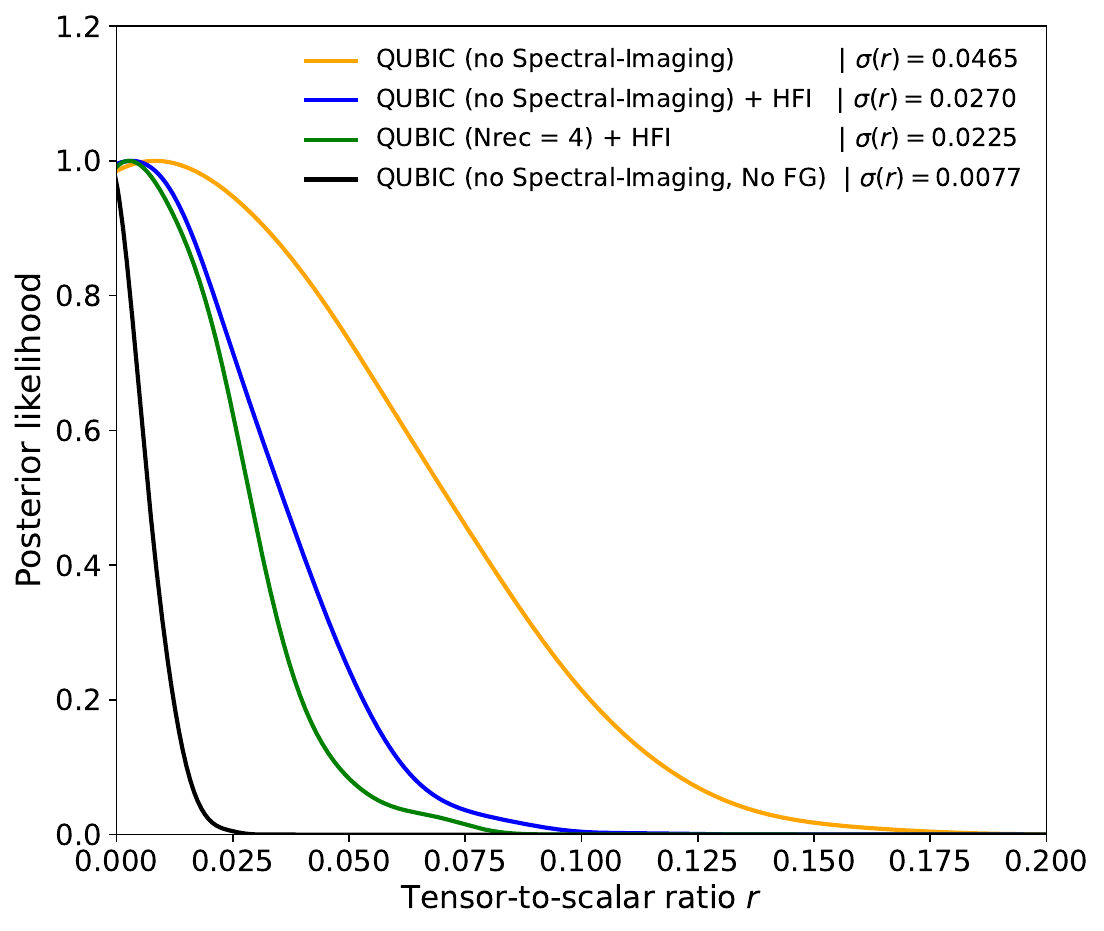}
        \caption{Posterior distribution on $r$ assuming CMB + thermal dust + noise. We assume no synchrotron contribution.}
    \label{fig:triangle_cosmo}
    \end{figure}

    We then performed the likelihood analysis in a Gaussian approximation~\citep{Hamimeche_2008} by comparing the reconstructed spectra with the model. We follow a similar procedure as in \citep{wolz2023simons}. As the QUBIC sky coverage does not allow to access low multipoles, we fix the reionization optical depth to $\tau = 0.054$~\citep{planck_2020}, and because of our half-degree angular resolution and sensitivity above $r=0.001$, we also fix the lensing amplitude to 1. We show our results in Fig.~\ref{fig:triangle_cosmo} for the final posterior likelihood on the tensor-to-scalar ratio $r$. We consider four cases:
    \begin{itemize}
        \item In the first case, we consider that primordial fluctuations are not contaminated by any foreground. Although unrealistic, this assumption allows us to have an estimation of the raw sensitivity of QUBIC on CMB polarization about $\sigma(r) = 0.0077$. Note that we found QUBIC is twice more sensitive to B-modes compared to the previous study in \cite{2020.QUBIC.PAPER1} mainly because we removed the edges effect on reconstructed frequency maps, and we consider the full covariance matrix during the $r$ reconstruction. 
        \item In the second case, we now introduce thermal dust contamination in the raw data. We use only QUBIC data to constrain the set of parameters $[ r, A_d, \alpha_d ]$ of the model in Eq.~\ref{eq:MCT_eq}. This procedure allows us to estimate a constraint $\sigma(r) = 0.0465$ on primordial B-modes using only QUBIC.
        \item In the third case, we use the same reconstructed frequency maps as in the previous case. We include in the likelihood procedure external data such as Planck maps to have stronger constraints on thermal dust. We use only the high-frequency instrument (HFI) of Planck because we have set the synchrotron amplitude to 0. This new dataset allows us to constrain the tensor-to-scalar ratio down to $\sigma(r) = 0.0270$.
        \item In the last case, we now apply the spectral-imaging on the same dataset to reconstruct twice the number of reconstructed frequency maps for QUBIC. An example of the bandpower noise covariance matrix is shown in Fig.~\ref{fig: corr_nrec4}, exhibiting positive and negative noise correlation between pairs of spectra. Using external data, we benefit from the increased spectral resolution of QUBIC to constrain spectral parameters, which impact the primordial B-modes reconstruction down to $\sigma(r) = 0.0225$. Although we operate in a simplified framework, the reduction of $\sigma(r)$ that we find in the case 4 compared to the previous one, provide evidence of the benefit when applying spectral imaging technique.
    \end{itemize}
    
    The method described here relies on several simplifying assumptions: stable atmosphere, relatively simplistic millimeter-wave sky, with dust characterized by a single spectral index over the sky, perfect knowledge of the synthesized beam and in general, no residual systematics. Despite these assumptions, it gives a good feeling of the efficiency of the map-making method presented in this article. This will be complemented by a more extensive study, currently ongoing, based on end-to-end simulations, which will give the actual forecasts for QUBIC, with and without sub-band splitting.
    
    \section{Conclusions}

    In this paper, we have demonstrated the novel capability of bolometric interferometry, specifically the QUBIC instrument, to perform spectral-imaging by generating multiple sub-frequency maps within each physical band. This advancement in spectral resolution holds promise for enhancing astrophysical component separation, particularly in scenarios where foregrounds exhibit complex SEDs beyond idealized models \citep{regnier2023identifying}. 
    
    Our approach relies on a sophisticated software-based method that leverages the frequency-dependent variation of the synthesized beam to achieve spectral-imaging. By formulating the problem as an inverse approach and minimizing a time-domain cost function, we accurately reconstruct frequency maps while addressing challenges such as edge effects and varying angular resolution across the instrument's bandwidth. External data, in this case taken from Planck observations, plays a crucial role in regularizing solutions and improving accuracy near map edges affected by peculiar beam characteristics of the QUBIC instrument.

    \begin{figure}
        \centering
        \includegraphics[width=0.5\linewidth]{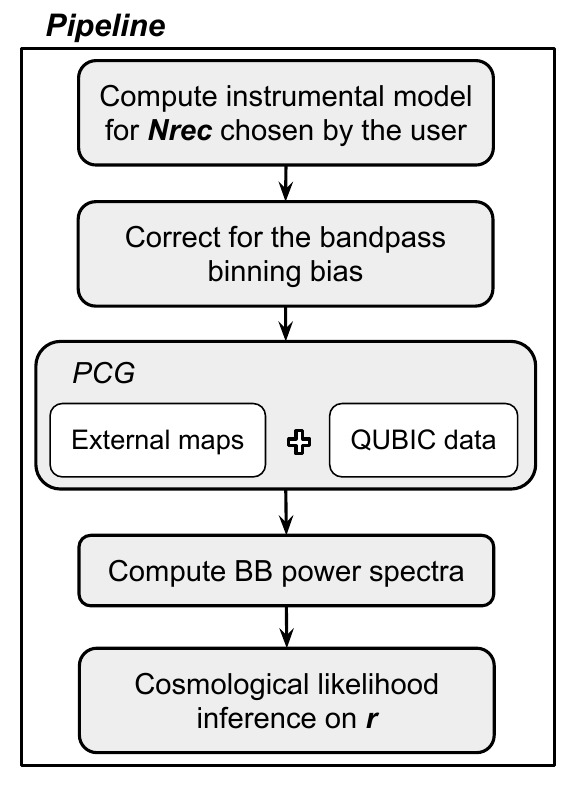}
        \caption{Scheme of all the steps of the analysis pipeline. From the top to the bottom are the first to the last steps. }
        \label{fig:resume}
    \end{figure}
    
    We have carried out simulations with different configurations to explore sky reconstruction and noise properties. We have identified noise anti-correlations between neighboring sub-bands at a significant level due to the spectral-imaging technique. Although these can, in principle, be accounted for in subsequent analyses (astrophysical components separation, power spectrum analysis, and cosmological parameters likelihood), this band-band noise correlation might induce significant algorithmic complications beyond the scope of this article. For the sake of simplicity, we have therefore achieved component separation and tensor-to-scalar ratio reconstruction using a single band reconstruction per physical band (thus, as a classical imager would do) and a classical cross-power-spectra analysis. We achieve a constraint on the tensor-to-scalar ratio of $\sigma(r) = 0.027$ after component separation and for three years at the nominal noise of QUBIC. We also performed spectral-imaging to the same dataset to increase the spectral resolution of the reconstructed maps, reaching a stronger constraint on $r$ with an uncertainty $\sigma(r) = 0.0225$. We have drawn a scheme in Fig.~\ref{fig:resume} to show all the different steps in the analysis pipeline. We start from the computation of the instrumental model, then we followed the iterative procedure to finally compute BB power spectra to infer the tensor-to-scalar ratio $r$ with some external datasets.
    
    The results we achieve in this study highlight how QUBIC, and more generally BI, provides a radically different manner of observing the CMB as well as the astrophysical foregrounds. Thanks to spectral-imaging, QUBIC can access detailed spectral information on the foregrounds within the instrumental bandwidth, providing a more robust assessment of their contamination. This is particularly required in the current phase of the primordial B-modes search, where foreground complexity appears as the biggest challenge.
    
    The algorithmic complexity induced by band-band noise correlations with spectral-imaging motivates the development of an alternative reconstruction technique called "components map-making". This new technique, aims at performing component separation and map-making in a single step, only modeling the noise in the time domain, therefore avoiding the complexity of band-band noise correlations. It is published in the separate CMM article~\citep{cmm}.
    
    Future developments will involve comparing the efficiency of our current frequency map-making approach with component map-making across various dust spectral behaviors to refine forecasts on the tensor-to-scalar ratio. Additionally, incorporating instrumental systematics into our inverse method approach promises to further enhance performance.

    \acknowledgments
    QUBIC is funded by the following agencies. France: ANR (Agence Nationale de la
    Recherche) contract ANR-22-CE31-0016, DIM-ACAV (Domaine d’Intérêt Majeur-Astronomie et Conditions d’Apparition de la Vie), CNRS/IN2P3 (Centre national de la recherche scientifique/Institut national de physique nucléaire et de physique des particules), CNRS/INSU (Centre national de la recherche scientifique/Institut national et al de sciences de l’univers). Italy: CNR/PNRA (Consiglio Nazionale delle Ricerche/Programma Nazionale Ricerche in
    Antartide) until 2016, INFN (Istituto Nazionale di Fisica Nucleare) since 2017.  Argentina: MINCyT (Ministerio de Ciencia, Tecnología e Innovación), CNEA (Comisión Nacional de Energía Atómica), CONICET (Consejo Nacional de Investigaciones Científicas y Técnicas).

    \bibliographystyle{JHEP}
    \bibliography{biblio}

    \appendix
    
    \section{Sample covariance calculation}
    \label{sec: appendixA}
    \newcommand{\xspec}[3]{\mathcal{C}_{#1}^{\nu_{#2} \times \nu_{#3}}}
    
    During the likelihood analysis, the noise band power covariance matrix is computed from noise-only simulations. The limited size of the observed sample compared to the true underlying distribution, due to the fact that we observe one realization of the sky over the infinite set of possibilities, causes an uncertainty on the observed data called ``cosmic variance'' or ``sample variance''. The sample variance, proportional to the signal, is accounted for by computing the covariance matrix between two cross-spectra. We express this quantity as:
    \begin{align}
        &\text{Cov}\left( \xspec{\ell}{i}{j},\, \xspec{\ell'}{k}{l} \right) = \left\langle \xspec{\ell}{i}{j} \xspec{\ell'}{k}{l}\right\rangle - \left\langle \xspec{\ell}{i}{j} \right\rangle \left\langle \xspec{\ell'}{k}{l} \right\rangle \nonumber \\
        &~~~~~~~~~~~~~~~~~~~~~~~~~~=\left( \frac{1}{2 \ell + 1} \right) \left( \frac{1}{2 \ell' + 1} \right) \sum_{m = - \ell}^{+ \ell} \sum_{m' = - \ell'}^{+ \ell'} \nonumber \\
        &~~~~~~~~~~~~~~~~~~~~~~~~~~~~~~~~~~~~~~~~~~~ \quad \left[ \left\langle a^{\nu_i}_{\ell m} a^{\nu_j}_{\ell m} a^{\nu_k}_{\ell' m'} a^{\nu_l}_{\ell' m'} \right\rangle \right. \nonumber \\
        &~~~~~~~~~~~~~~~~~~~~~~~~~~~~~~~~~~~~~~~~~~~ \quad \quad- \left. \left\langle a^{\nu_i}_{\ell m} a^{\nu_j}_{\ell m} \right\rangle \left\langle a^{\nu_k}_{\ell' m'} a^{\nu_l}_{\ell' m'} \right\rangle \right]. \label{eq: cov_appendix}
    \end{align}
    
    We can use Wick's identity for Gaussian random variables, giving:
    \begin{align*}
        \left\langle a^{\nu_i}_{\ell m} a^{\nu_j}_{\ell m} a^{\nu_k}_{\ell' m'} a^{\nu_l}_{\ell' m'} \right\rangle = 
        &\left\langle a^{\nu_i}_{\ell m} a^{\nu_j}_{\ell m} \right\rangle \left\langle a^{\nu_k}_{\ell' m'} a^{\nu_l}_{\ell' m'} \right\rangle \\
        &~~~~~~+ \left\langle a^{\nu_i}_{\ell m} a^{\nu_k}_{\ell' m'} \right\rangle \left\langle a^{\nu_j}_{\ell m} a^{\nu_l}_{\ell' m'} \right\rangle \\
        &~~~~~~~~~~~~~+ \left\langle a^{\nu_i}_{\ell m} a^{\nu_l}_{\ell' m'} \right\rangle \left\langle a^{\nu_j}_{\ell m} a^{\nu_k}_{\ell' m'} \right\rangle.
    \end{align*}
    
    Replacing this expression in Eq.~\ref{eq: cov_appendix}, we have:
    \begin{align*}
        &\text{Cov}\left( \xspec{\ell}{i}{j},\, \xspec{\ell'}{k}{l} \right) = \left( \frac{1}{2 \ell + 1} \right) \left( \frac{1}{2 \ell' + 1} \right) \sum_{m = - \ell}^{+ \ell} \sum_{m' = - \ell'}^{+ \ell'} \\
        &~~~~~~~~~~~~~~~~~~~~~~~~~~~~~~~~~~~~~~~~~~~~~~~~\quad \quad \left[ \left\langle a^{\nu_i}_{\ell m} a^{\nu_k}_{\ell' m'} \right\rangle \left\langle a^{\nu_j}_{\ell m} a^{\nu_l}_{\ell' m'} \right\rangle \right. \\
        &~~~~~~~~~~~~~~~~~~~~~~~~~~~~~~~~~~~~~~~~~~~~~~~~\quad \quad + \left. \left\langle a^{\nu_i}_{\ell m} a^{\nu_l}_{\ell' m'} \right\rangle \left\langle a^{\nu_j}_{\ell m} a^{\nu_k}_{\ell' m'} \right\rangle 
        \right]
    \end{align*}
    
    \begin{align*}
        &\text{Cov}\left( \xspec{\ell}{i}{j},\, \xspec{\ell'}{k}{l} \right) = \left( \frac{1}{2 \ell + 1} \right) \left( \frac{1}{2 \ell' + 1} \right) \sum_{m = - \ell}^{+ \ell} \sum_{m' = - \ell'}^{+ \ell'} \delta_{\ell \ell'} \delta_{m m'} \\
        &~~~~~~~~~~~~~~~~~~~~~~~~~~~~~~~~~~~~~~~~~~~~~~~~\quad \quad\left[ \xspec{\ell}{i}{k} \xspec{\ell}{j}{l} + \xspec{\ell}{i}{l} \xspec{\ell}{j}{k} \right].
    \end{align*}
    
    Observing that $\sum_{m = - \ell}^{+ \ell} \sum_{m' = - \ell'}^{+ \ell'} \delta_{\ell \ell'} \delta_{m m'} = 2 \ell' + 1$, we finally have:
    \begin{equation*}
        \text{Cov}\left( \xspec{\ell}{i}{j},\, \xspec{\ell'}{k}{l} \right) = \frac{1}{2 \ell + 1} \left[ \xspec{\ell}{i}{k} \xspec{\ell}{j}{l} + \xspec{\ell}{i}{l} \xspec{\ell}{j}{k} \right],
    \end{equation*}
    where the element $f_{\text{sky}}$ and $\Delta_{\ell}$ are manually added to account for the cut sky and binned spectra (see equation~\ref{eq: samplevariance}).

    \section{Resolution of the final maps without convolutions during reconstruction} \label{sec: appendixB}
    
    As explained in section~\ref{sect:angres}, we do the reconstruction with a simulated TOD written as:
    \begin{equation}
        \vec{\tilde{d}} = \sum_{i = 1}^{N_{\text{rec}}} \left( \sum_{j = 1}^{f_{\text{sub}}}
        \Delta\nu_{ij} \mathcal{H}_{\nu_{ij}} \mathcal{C}_{K_{ij}} \right) \cdot \vec{\tilde{s}}_{\nu_i}.
    \end{equation}
    This is computationally very expensive because of the convolution operators $\mathcal{C}_{K_{ij}}$. So we want to explore the possibility of not doing the convolutions during the reconstruction. The real TOD is still generated through convoluted maps (see Eq.~\ref{eq:data_angres}), so the PCG will try to fit a map $\vec{\Tilde{s}}_{\nu_i}^{\sigma_i}$ at some resolution $\sigma_i$, common for the $f_\text{sub}$ frequencies of the sub-band $i$. The simulated TOD is then:
    \begin{equation}
        \vec{\Tilde{d}} = \sum_{i=1}^{N_\text{rec}} \sum_{j=1}^{f_{\text{sub}}} \Delta\nu_{ij} \mathcal{H}_{\nu_{ij}} \vec{\Tilde{s}}_{\nu_i}^{\,\sigma_i}.
    \end{equation}
    We want to determine $\sigma_i$ and $\nu_i$. The PCG fits the simulated TOD to the real TOD, so we get, for each of the $N_\text{rec}$ sub-bands, the equality:
    \begin{equation}
        \sum_{j=1}^{f_\text{sub}} \Delta\nu_{ij} \mathcal{H}_{\nu_{ij}} \vec{\Tilde{s}}_{\nu_i}^{\,\sigma_i} \approx \sum_{j=1}^{f_\text{sub}} \Delta\nu_{ij} \mathcal{H}_{\nu_{ij}} \mathcal{C}_{\sigma_{\nu_{ij}}} \vec{s}_{\nu_{ij}}^{\,\infty},
    \end{equation}
    \begin{equation}
        \vec{\Tilde{s}}_{\nu_i}^{\,\sigma_i} \approx \left( \sum_{j=1}^{f_\text{sub}} \Delta\nu_{ij} \mathcal{H}_{\nu_{ij}} \right)^{-1} \cdot \sum_{j=1}^{f_\text{sub}} \Delta\nu_{ij} \mathcal{H}_{\nu_{ij}} \mathcal{C}_{\sigma_{\nu_{ij}}} \vec{s}_{\nu_{ij}}^{\,\infty}.
    \end{equation}
    This means that the reconstructed map $\vec{\Tilde{s}}_{\nu_i}^{\,\sigma_i}$ is the average of the maps $\mathcal{C}_{\sigma_{\nu_{ij}}} \vec{s}_{\nu_{ij}}^{\,\infty}$ weighted by the operators $\Delta\nu_{ij} \mathcal{H}_{\nu_{ij}}$. It's not very practical to weight an average with operators, so we introduce the scalars $h_{\nu_{ij}}$:
    \begin{equation}
        h_{\nu_{ij}} = \mathcal{H}_{\nu_{ij}} \cdot \mathcal{I},
    \end{equation}
    which represent the scale of the operators $\mathcal{H}_{\nu_{ij}}$ and $\mathcal{I}$ is a uniform sky set to one everywhere. Then we can write the reconstructed map as:
    \begin{equation}
        \vec{\Tilde{s}}_{\nu_i}^{\,\sigma_i} \approx \dfrac{\sum_{j=1}^{f_\text{sub}} \Delta\nu_{ij} h_{\nu_{ij}} \mathcal{C}_{\sigma_{\nu_{ij}}} \vec{s}_{\nu_{ij}}^{\,\infty}}{\sum_{j=1}^{f_\text{sub}} \Delta\nu_{ij} h_{\nu_{ij}}} = \left\langle \mathcal{C}_{\sigma_{\nu_{ij}}} \vec{s}_{\nu_{ij}}^{\,\infty} \right\rangle_j,
    \end{equation}
    where we introduced the notation $\langle . \rangle_j$ which is the average over $j$ weighted by $\Delta\nu_{ij} h_{\nu_{ij}}$.
    
    We can decompose the sky as the sum of the CMB and the dust (we neglect the synchrotron, see section~\ref{sec: analysis principle}): $\vec{s}_{\nu_{ij}}^{\,\infty} = \vec{s}_{\text{CMB}}^\infty + f_d^{\,\beta}(\nu_{ij}) \cdot \vec{s}_{\text{dust}}^\infty$. Then we have:
    \begin{equation}
        \left\langle \mathcal{C}_{\sigma_{\nu_{ij}}} \vec{s}_{\nu_{ij}}^{\,\infty} \right\rangle_j = \left\langle \mathcal{C}_{\sigma_{\nu_{ij}}} \right\rangle_j \cdot \vec{s}_{\text{CMB}}^\infty + \left\langle f_d^{\,\beta}(\nu_{ij}) \mathcal{C}_{\sigma_{\nu_{ij}}} \right\rangle_j \cdot \vec{s}_{\text{dust}}^\infty.
    \end{equation}
    Let us concentrate on the dust term. The average frequency $\nu_i$ that we are trying to determine won't depend on the convolution operator $C_{\sigma_{\nu_{ij}}}$ as it is normalized. From this, we understand that $f_d^{\,\beta}(\nu_i) = \langle f_d^{\,\beta}(\nu_{ij}) \rangle_j$, which gives a precise definition for $\nu_i$:
    \begin{equation}
        \nu_i = f_d^{\,\beta\, -1}\left(\left\langle f_d^{\,\beta}(\nu_{ij}) \right\rangle_j\right),
    \end{equation}
    with $f_d^{\,\beta\, -1}$ the inverse function of $f_d^{\,\beta}$.
    
    We can now write:
    \begin{equation}
        \left\langle \mathcal{C}_{\sigma_{\nu_{ij}}} \vec{s}_{\nu_{ij}}^{\,\infty} \right\rangle_j = \left\langle \mathcal{C}_{\sigma_{\nu_{ij}}} \right\rangle_j \cdot \vec{s}_{\text{CMB}}^\infty + \left\langle \frac{f_d^{\,\beta}(\nu_{ij})}{f_d^{\,\beta}(\nu_i)} \mathcal{C}_{\sigma_{\nu_{ij}}} \right\rangle_j \cdot f_d^{\,\beta}(\nu_i) \vec{s}_{\text{dust}}^\infty.
    \end{equation}
    We have:
    \begin{equation}
        \left\langle \frac{f_d^{\,\beta}(\nu_{ij})}{f_d^{\,\beta}(\nu_i)} \mathcal{C}_{\sigma_{\nu_{ij}}} \right\rangle_j = \frac{\sum_{j=1}^{f_\text{sub}} \Delta\nu_{ij} h_{\nu_{ij}} f_d^{\,\beta}(\nu_{ij}) \mathcal{C}_{\sigma_{\nu_{ij}}}}{\sum_{j=1}^{f_\text{sub}} \Delta\nu_{ij} h_{\nu_{ij}} f_d^{\,\beta}(\nu_{ij})} = \left[\mathcal{C}_{\sigma_{\nu_{ij}}}\right]_j,
    \end{equation}
    where we introduced the notation $[ . ]_j$ which is the average over $j$ weighted by $\Delta\nu_{ij} h_{\nu_{ij}} f_d^{\,\beta}(\nu_{ij})$.
    
    Dust is dominant compared to CMB, so the final resolution of the map is the one of the dust map. We can approximate $\left[\mathcal{C}_{\sigma_{\nu_{ij}}}\right]_j$ by a convolution with a Gaussian function of width:
    \begin{equation}
        \sigma_i = \left[\sigma_{\nu_{ij}}\right]_j.
    \end{equation}
    This is the resolution of the map of frequency $\nu_i$ after the reconstruction without the convolutions during the process.

    \section{View of the radial profile} \label{sec: AppendixC}

    \begin{figure}[h!]
        \centering
    	\includegraphics[scale = 0.5]{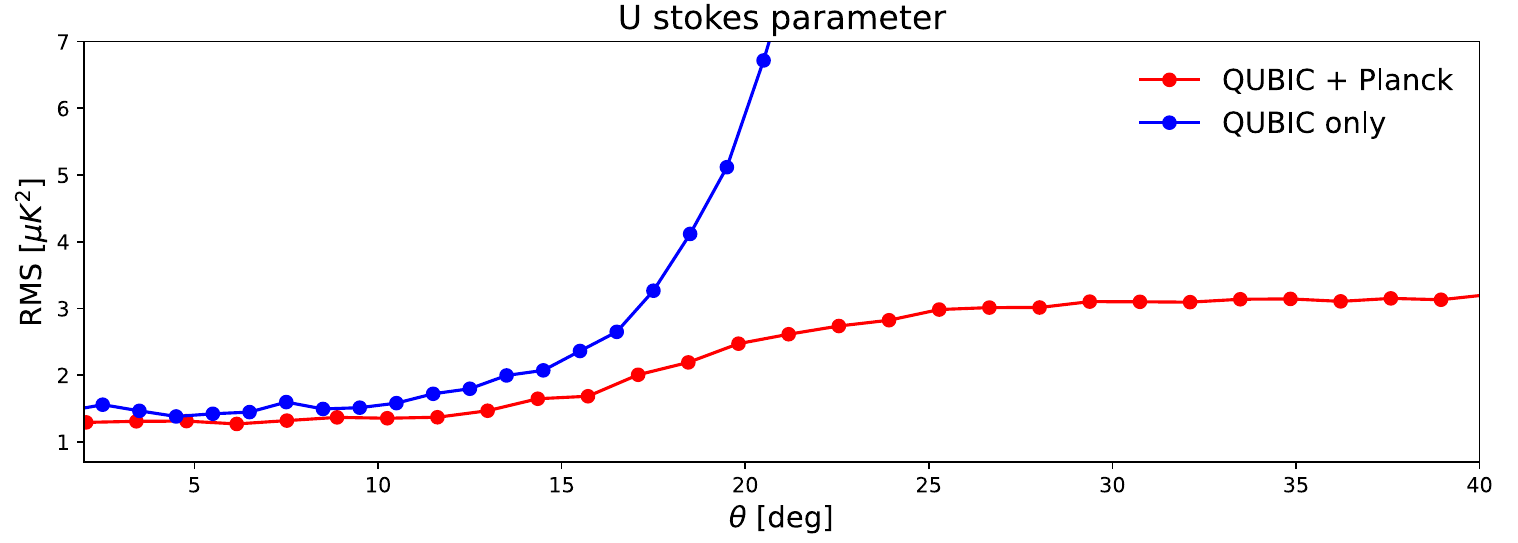}
    	\caption{Broadband case - Profile of the residual RMS for Q (\textit{top}) and U \textit{(bottom)} Stokes parameter. The blue line shows the RMS using QUBIC only and the red line shows the RMS when using  Planck data to regularize the edge effects.}
    	\label{fig:profile_broadband}
    \end{figure}
    
    To have a better visualization of the problematic edges effect, we show the radial profile obtained in two cases. The first (in blue) show the original reconstruction with edge effects and the second (in red), shows the reconstruction with the regularization. Fig.~\ref{fig:profile_broadband} shows the residual RMS profile as a function of angular distance from the center of the QUBIC patch in $\text{(RA, DEC)} = (0^{\circ}, -57^{\circ})$ for the U stokes parameter. The same curve is obtained for the Q stokes parameter. The improvement provided by the inclusion of the Planck for regularization is obvious and allows achieving a rather flat residual RMS profile.

\end{document}